\documentclass[layout=standard]{achemso}

\usepackage[version=3]{mhchem} 
\usepackage{graphicx} 
\usepackage{bm} 
\usepackage{longtable}
\usepackage{multirow}

\newcommand{\bra}[1]{\langle #1 |}
\newcommand{\ket}[1]{| #1 \rangle}
\newcommand{\crea}[2]{#1^{\dagger}_{#2}}
\newcommand{\anni}[2]{#1_{#2}}

\author{Felipe Herrera}
\email{fherreraurbina@fas.harvard.edu}
\affiliation{Department of Chemistry and Chemical Biology, Harvard University, Cambridge, USA 02138}

\author{Borja Peropadre}
\affiliation{Department of Chemistry and Chemical Biology, Harvard University, Cambridge, USA 02138}

\author{Leonardo A. Pachon}
\affiliation{Department of Chemistry and Chemical Biology, Harvard University, Cambridge, USA 02138}
\altaffiliation{Grupo de F\'isica At\'omica y Molecular, Instituto de F\'{\i}sica,  Facultad de Ciencias Exactas y Naturales, Universidad de Antioquia UdeA; Calle 70 No. 52-21, Medell\'in, Colombia.}

\author{Semion K. Saikin}
\affiliation{Department of Chemistry and Chemical Biology, Harvard University, Cambridge, USA 02138}
\altaffiliation{Institute of Physics, Kazan Federal University, 18 Kremlevskaya Street, Kazan, 420008, Russian Federation}

\author{Al\'an Aspuru-Guzik}
\email{aspuru@chemistry.harvard.edu}
\affiliation{Department of Chemistry and Chemical Biology, Harvard University, Cambridge, USA 02138}
\date{\today}

\title{Quantum nonlinear optics with polar J-aggregates in microcavities}

\abbreviations{EIT, cQED}

\begin{document}

\begin{tocentry}
\centerline{\includegraphics[height=3.5cm]{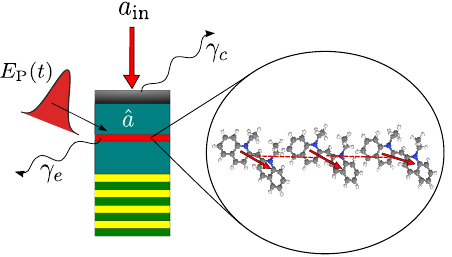}}
\end{tocentry}

\newpage
\begin{abstract}
We show that an ensemble of organic dye molecules with permanent electric dipole moments embedded in a microcavity can lead to strong optical nonlinearities at the single photon level. The strong long-range electrostatic interaction between chromophores due to their permanent dipoles introduces the desired nonlinearity of the light-matter coupling in the microcavity. 
We obtain the absorption spectra of a weak probe field under the influence of strong exciton-photon coupling with the cavity field. 
Using realistic parameters, we demonstrate that a single cavity photon can significantly modify the absorptive and dispersive response of 
the medium to a probe photon at a different frequency. 
Finally, we show that the system is in the regime of cavity-induced transparency with a broad transparency window for dye dimers. We illustrate our findings using pseudoisocyanine chloride (PIC) J-aggregates in currently-available optical microcavities.
\end{abstract}

\newpage
J-aggregates \cite{Jelley1936,Scheibe1936,Kobayashi-book,Agranovich-book,Saikin:2013} are arrays of dye molecules with large dipole moments that exhibit strong intermolecular electrostatic interaction, giving rise to collective effects in their coupling with electromagnetic fields. The specific set of linear and nonlinear optical properties of J-aggregates has stimulated a resurgence of interest in them for applications in modern photonics.
A large linear absorption cross section combined with a narrow line width \cite{Wurthner2011} at room temperatures make J-aggregates attractive for the design of optical processing devices operating at low light levels. J-aggregates can be readily coupled to solid state photonic \cite{Lidzey:1998, Coles2014} and plasmonic \cite{Bellessa:2004,Vasa2013,Zengin:2013} structures extending the conventional photonics to sub-diffraction length scales \cite{Saikin:2013}. An illustrative example of molecular-based photon processing structures can be found in nature, where photosynthetic organisms use molecular aggregates to collect light and deliver the photon energy on the scale of tens of nanometers \cite{Scholes2011}. 

Moderately strong laser fields are commonly used in free-space to observe coherent optical phenomena in atomic gases and a few solid-state systems characterized by long dephasing times exceeding milliseconds at room temperature~\cite{Fleischhauer:2005,Shore:2008,Lee:2011}. Solid-state semiconducting materials have much shorter electronic coherence times on the order of hundreds of femtoseconds, which greatly increases the laser intensity required to induce coherent optical phenomena in free space. For instance, in order to observe electromagnetically-induced transparency (EIT) using inorganic quantum dots with terahertz dephasing rates, the required control laser intensity should be on the order of tens of MW/cm$^2$\cite{Houmark:2009}. The same applies for organic materials, including J-aggregates. Such high intensities can optically damage an organic medium~\cite{Akselrod:2010}. It is therefore necessary to replace the control lasers by the strong electric field per photon achievable in photonic structures~\cite{Saikin:2013}, in order to observe coherent optical response with organic matter at room temperature.  

Experimental progress in the fabrication of organic optical microcavities has demonstrated the ability to strongly couple an ensemble of organic chromophores with the confined electromagnetic field of a cavity mode at room temperature \cite{Lidzey:1998,Schouwink:2001,Tischler:2005,Kena-Cohen:2010,Kena-Cohen:2008,Bittner:2012}, via the emergence of polariton modes in the cavity transmission spectra. The strong coupling of organic ensembles with plasmonic modes has also been demonstrated~\cite{Bellessa:2004,Fogang:2008, Ditinger:2005,Wurtz:2007,Sugawara:2006,Saikin:2013}. Moreover, the regime of ultrastrong coupling with organic molecules is now within reach, where the light-matter interaction energy reaches a significant fraction of the associated transition frequency\cite{Schwartz:2011}. These experimental advances enable the possibility of understanding and possibly manipulating the excited state dynamics of molecular aggregates using a small number of photons.

In this Letter, we address the question whether collective multi-exciton states in J-aggregates can be exploited for the coherent control of confined optical fields in photonic structures. In order to achieve this, we extend the nonlinear exciton equation (NEE) formalism~\cite{Chernyak:1998,Mukamel:2004} to account for the non-perturbative coupling of the medium to a confined optical field. 
As an example, we consider an ensemble of one-dimensional polar J-aggregate domains embedded in an optical microcavity as a non-linear optical material with a substantial response at low light levels. 
We demonstrate that by exploiting the strong dipole-dipole interaction between individual chromophores due to their permanent dipoles, plus the strong collective coupling of a molecular aggregate with the cavity field, it is possible to perform light-by-light switching at the single-photon level. Specifically, we show that the presence of a single photon at the cavity frequency can modify absorptive and dispersive response of the organic medium to a weak external probe at a different frequency. The intermolecular electronic coupling between chromophores is responsible for establishing the required anharmonicities in the material spectrum, and the large electric field per photon of the confined cavity mode reduces the number of control photons required to achieve an observable switching effect. 

To describe the evolution of the medium polarization $\mathbf{P}(t)$, we employ a quantum Langevin formalism. The key features of our model are ({\it i}) the strong coupling of the molecular ensemble with the cavity field, and ({\it ii}) the intermolecular resonant energy tranfer via transition dipoles, known as F\"orster coupling $J_{ij}$, in addition to diagonal dipole-dipole interaction $U_{ij}$ via permanent dipoles. Additionally we consider chromophore relaxation due to spontaneous emission outside the confined cavity mode, coupling of the chromophores to a phonon bath, and inhomogenous broadening due to static disorder in chromophore transition energies. The evolution of an observable $O$ in the Heisenberg picture is given by $d  O/dt = -i[O,\mathcal{H}_{\rm S}+\mathcal{H}_{\rm SB}]$ (we use $\hbar=1$ throughout), where the Hamiltonian $\mathcal{H}_{\rm S}$ describes the coherent evolution of the system degrees of freedom, and $\mathcal{H}_{SB}$ the interaction of the system with the environment. More specifically, 
$\mathcal{H}_{\rm S}$ describes the interaction of a single planar J-aggregate containing $N$ chromophores with the electromagnetic field of a single cavity mode at frequency $\omega_c$ as well as a probe 
field at frequency $\omega_p$, and can be partitioned as
\begin{eqnarray}
\mathcal{H}_{\rm S} = H_1+H_2+H_3.
\label{eq:system H}
\end{eqnarray}
The first term describes a single effective molecular aggregate, as defined in the Supporting Information (SI), in the one-exciton eigenbasis as
\begin{eqnarray}
 H_1 = \sum_k \omega_k B_k^\dagger B_k +\sum_{kp}U_{kp}\crea{B}{k}\crea{B}{p}\anni{B}{k}\anni{B}{p}.
 \label{eq:exciton}
\end{eqnarray}
The bosonic operator $\anni{B}{k}$ annihilates an exciton in $k$-th mode, with $k=\{1,2,\ldots, N\}$. We assume the aggregate is a collection of two-level chromophores, with ground state $\ket{g}$ and excited state $\ket{e}$ having a site dependent transition energy $\varepsilon_i = \omega_e+ d_i$, where $d_i$ is a small random shift from the gas-phase transition frequency $\omega_e$ that models structural or so-called static disorder \cite{Knapp:1984,Knoester:1993}. 
The first term in eq. (\ref{eq:exciton}), is the diagonal form of the site-basis Frenkel exciton Hamiltonian $H_0 = \sum_i \varepsilon_i\crea{B}{i}\anni{B}{i} +\sum_{ij}J_{ij}\crea{B}{i}\anni{B}{j}$. In the point dipole approximation, the exchange coupling energy is $J_{ij} = (1-3\cos^2\Theta_{ij})d_{eg}^2/r_{ij}^3$, where $\Theta_{ij}$ is the angle between the transition dipole moments of molecules $i$ 
and $j$ (assumed parallel) and the intermolecular separation vector $\mathbf{r}_{ij}=r_{ij}\hat{\bf r}_{ij}$.  The second term in eq. (\ref{eq:exciton}) describes the interaction between two exciton eigenstates due to long-range Coulomb forces between the permanent dipoles of the chromophores. Here we assume a simplified form of the scattering potential $U_{kp} = \sum_{ij}U_{ij}|c_{ik}|^2|c_{jp}|^2$, where $c_{ik}$ is an element of the unitary transformation $\anni{B}{i} = \sum_k c_{ik}\anni{B}{k}$. The interaction energy between sites is $U_{ij} = (1-3\cos^2\Theta_{ij})(\Delta d)^2/r_{ij}^3$, where $\Delta d = d_e-d_g$ is the change in permanent dipole moment upon excitation of the chromophores \cite{Spano:1991}. For homogeneous aggregates, large values of $U_{12}$ can lead to the formation of biexcitons with a binding energy proportional to $U_{12}$\cite{Spano:1991}. In this work we simplify the two-exciton problem by assuming that the leading effect of the potential $U_{kp}$ is to red-shift or blue-shift 
the two-exciton band with respect to the non-interacting case, for attractive or repulsive interactions, respectively. For simplicity, we take two-exciton eigenstates as direct products of single-exciton states. 

The second term $H_2 = \omega_c a^\dagger a +\omega_p\mathcal{E}^\dagger\mathcal{E}$ in eq. (\ref{eq:system H}) is the free Hamiltonian for the cavity and probe fields, and the third term describes light-matter interaction as
\begin{eqnarray}
H_3 &=& i\sum_{k}g_k(t)(\mathcal{E}^\dagger B_k - B^\dagger_k \mathcal{E})+i\sum_{kq}D_{k,kq}(a^\dagger\crea{B}{k}\anni{B}{k}\anni{B}{q} - \crea{B}{q}\crea{B}{k}\anni{B}{k}a),
\end{eqnarray}
%
where $g_k(t)=\sqrt{N_A}(\vec{\mu}_k\cdot\mathbf{e}_p) {E}_p(t)$ is proportional to the single-exciton transition dipole moment $\vec{\mu}_k=\langle k|\vec{\mu}|g\rangle$ and $D_{k,kq}=\sqrt{N_A}(\vec{\mu}_{k,kq}\cdot \mathbf{e}_c)\mathcal{E}_c$ is proportional to the one-to-two exciton transition dipole moment $\mu_{k,kq}=\langle k|\vec{\mu}|kq\rangle$. The organic medium typically consists of an ensemble of aggregate domains\cite{Kato:1999,Kobayashi-book}. Within each domain the intermolecular interactions are much stronger than between domains. For simplicity, we idealize the medium by assuming that each domain contains a single one-dimensional aggregate, and all domains are identical. $N_A$ is the number of aggregates in the medium (details in the SI). ${E}_p(t)$ and $\mathbf{e}_p$ are the electric field envelope and polarization of the probe. $\mathcal{E}_c$ is the electric field per cavity photon and $\mathbf{e}_c$ its polarization. The probe and cavity polarizations are assumed to be collinear. 
\begin{figure}[t]
 \includegraphics[width=0.45\textwidth]{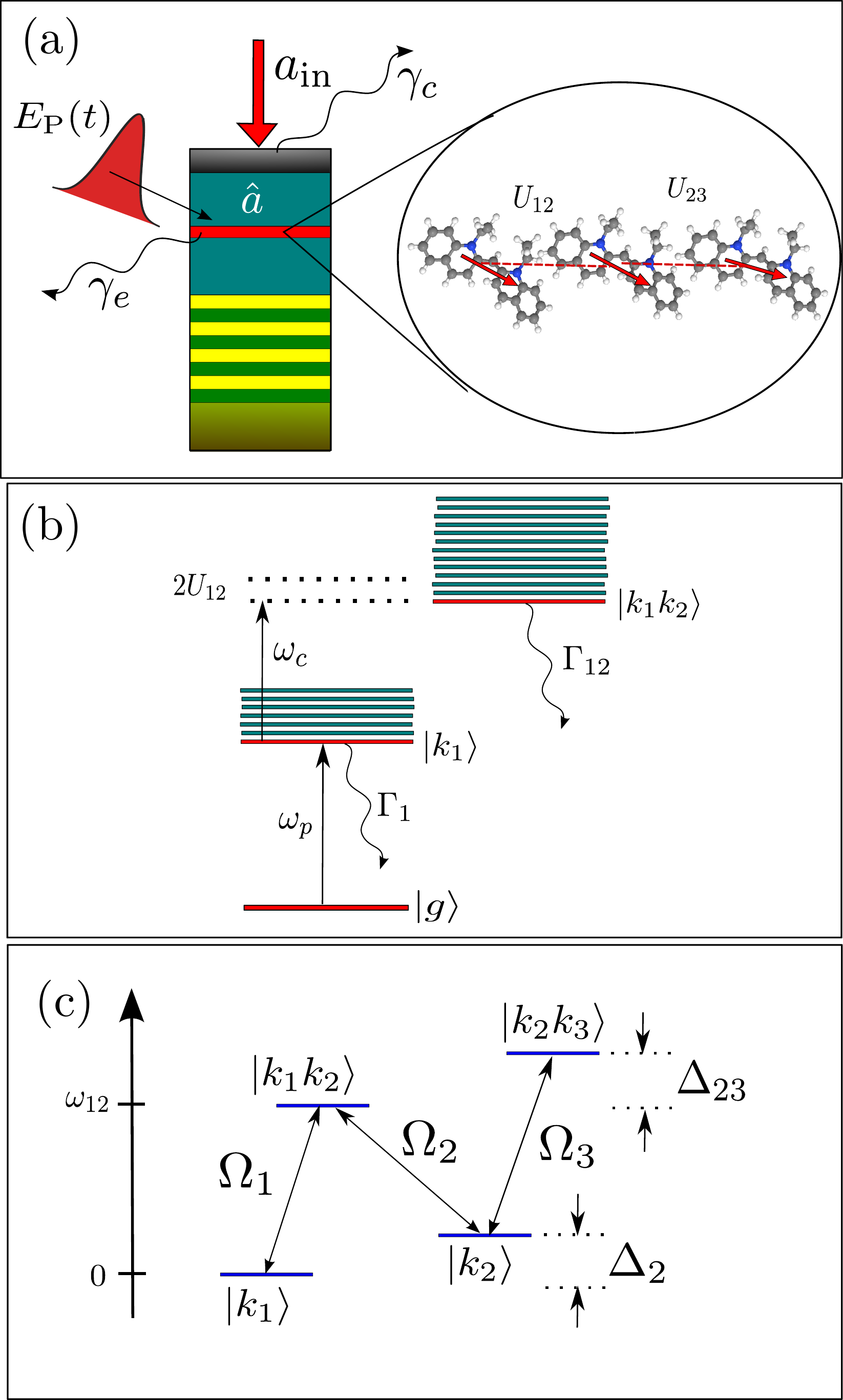}
 \caption{{\bf Panel (a)}: Illustration of an optical microcavity containing an ensemble of two-dimensional J-aggregates. The cavity mode $a_c$ is driven by a weak input field $a_{\rm in}$ and decays through the semi-reflecting mirror at the rate $\gamma_c$. A weak probe field at frequency $\omega_p>\omega_c$ couples directly to the organic chromophores. Individual molecules decay into external modes with a rate $\gamma_e$. Dipole-dipole interactions between individual chromophores in each aggregate modify the single-molecule response of the medium to the cavity and probe fields. {\bf Panel (b)}: Energy spectrum of an individual aggregate showing the one-exciton and two-exciton bands (bandwidths not on scale). The cavity field drives all the allowed coherences between states $\ket{k_i}$ and $\ket{k_ik_j}$ and the weak probe removes population from the ground state $\ket{g}$. The transition frequency between $\ket{k_i}$ and $\ket{k_ik_j}$ is shifted by the interaction energy $\sim U_{ij}$ with respect to the 
non-interacting case. {\bf Panel (c)}: Effective four-level system interacting coupled by the cavity with Rabi frequencies $\Omega_2>\Omega_1>\Omega_3$. The cavity frequency $\omega_c$ is assumed to be near resonance with the transition $\ket{k_1}\rightarrow\ket{k_2}$. This model is used in eq. (\ref{eq:effective H4}) to describe the probe absorption at frequency $\omega_p<\omega_c$.}
 \label{fig:driven cavity}
\end{figure}

The system-bath interaction is partitioned as $\mathcal{H}_{\rm SB}=H_{\rm ex}+H_{\rm cav}$, where $H_{\rm cav}$ describes the decay of the cavity mode through the mirror of a one-sided microcavity, which corresponds to a typical experimental setup\cite{Akselrod:2010}. The term $H_{\rm ex}$ describes the radiative decay of excitons into electromagnetic modes outside the cavity in addition to dephasing of excitons via interactions with phonons. The specific form $\mathcal{H}_{\rm SB}$ and the relaxation tensors for system observables used in this work are given in the SI. In Fig. \ref{fig:driven cavity} we illustrate the system under consideration and the spectrum including the two lowest exciton bands.

We are interested in the polarization $\mathbf{P}(t)$ of the medium, induced by the weak coherent probe field $\mathcal{E}$, with frequency $\omega_p$.
The medium polarization at frequency $\omega_p$ is given by 
\begin{eqnarray}
\mathbf{P}(t) = \sum_k \mathbf{\mu_k}\left\{\langle B_k(t)\rangle+\langle B_k^\dagger(t)\rangle\right\}.
\end{eqnarray}
We therefore need to solve the quantum Langevin equation for the exciton coherence $\langle B_k(t\rightarrow\infty)\rangle$ in the steady state. 
The nonlinearity in the system Hamiltonian $H_1$ couples the observable $B_k$ with an infinite hierarchy of equations of motion involving powers of the material operators $B_k$ and $B_k^\dagger$. Since we are interested in the interaction of the medium with at most one probe photon and one cavity photon on average, we invoke a Dynamics-Controlled Truncation scheme (DCT)\cite{Portolan:2008} to truncate the hierarchy at third order, thus neglecting correlation functions involving products of four or more excitonic variables. Given the small excitation density generated by the weak probe, we also assume that the ground state population remains near unity at all times, ignoring contributions from density terms such as $\langle B_k^\dagger B_k\rangle$ in the equations of motion. Moreover, we assume a semiclassical approximation for the cavity field, which amounts to factorizing the correlation functions involving producs of cavity and material variables. The dynamics of the medium polarization is thus governed by the equations 
\begin{eqnarray}\label{eq:B eom}
 \frac{d}{dt}\langle {B}_k\rangle &=& (-i\omega_k-\Gamma_k/2)\langle B_k\rangle  - g_k\langle \mathcal{E}\rangle +\sum_{pq} D_{k,pq} \langle a^\dagger\rangle \langle B_pB_{q}\rangle, \\
 \frac{d}{dt}\langle B_pB_q\rangle &=& [-i(\omega_p+\omega_q+2U_{pq})-\Gamma_{pq}]\langle B_pB_q\rangle - 2\sum_k D_{k,pq} \langle B_k\rangle \langle a\rangle - \left(g_p \langle B_q\rangle +g_q\langle B_p\rangle \right)\langle \mathcal{E}\rangle.\label{eq:BB eom}\nonumber\\
\end{eqnarray}
%
%
 %
Assuming that the phonon and the photon baths are Markovian, the Langevin noise terms in the equations of motion do not contribute to the evolution of the expectation values $\langle B_k\rangle$ and $\langle a\rangle$ \cite{Gardiner-book}. For simplicity, we have also neglected the effect of Langevin noise terms in the two-point and three-point correlation functions. We assume the cavity oscillator is weakly driven by a coherent field with $\langle a_{\rm in}\rangle>0$ in the underdamped regime. We have ignored contributions of three-point correlation functions of the form $\langle B_p^\dagger B_qB_k\rangle$, representing one-to-two exciton coherences. These coherence can be shown to remain negligible unless the ground state is depleated by the probe field beyond the perturbative regime. The cavity field couples directly to the $\langle B_p^\dagger B_qB_k\rangle$ in the Langevin equations (see SI). Therefore, in the perturbative regime with respect to the probe field, the cavity amplitude $\langle a(t)
\rangle$ evolves as if the cavity was empty.

Despite the number of simplifications made in the derivation of equations (\ref{eq:B eom}) and (\ref{eq:BB eom}), we note that they have the same structure as the nonlinear exciton equations (NEE) that Chernyak {\it et al.}\cite{Chernyak:1998,Mukamel:2004} derived by taking into account the non-boson commutation of exciton operators, density terms, and inelastic exciton-exciton scattering. Therefore, based on several previous studies of nonlinear optical spectroscopy using the NEE in the regime of perturbative light-matter interaction\cite{Mukamel:2004}, 
we expect our model to provide an accurate qualitative description of the nonlinear response of molecular aggregates in microcavities where the coupling to the cavity mode is non-perturbative. 

We are interested in the steady state response of the system to the probe field, for timescales long compared with the exciton and cavity lifetimes. Moreover, we assume the cavity field decays at a rate slower than the exciton coherence decay ($1/\Gamma_k\sim 10^2$ fs), which does not require very high-Q cavities at room temperature~\cite{Akselrod:2010}. This separation of timescales allows us to solve eqs. (\ref{eq:B eom})-(\ref{eq:BB eom}) taking the cavity and probe amplitudes as constants. Another important assumption in our model is the resolution of frequencies in the system. We require the detuning of the probe field from the cavity field $\delta = \omega_c-\omega_p$ to be larger than the exciton linewidth. We also want the cavity to interact resonantly with only a subset of transitions from the one-exciton band to the two-exciton band, so that only the weak probe field can (perturbately) create excitations in the medium when resonant with a one-exciton state. In order for the cavity not to generate excitations in the medium when resonantly driven by an external input field $\langle a_{\rm in}\rangle$, we require $\delta<0$ and $U_{kp}<0$ with $|\delta| \sim {\rm max}\{ |U_{kp}|\}\gg {\rm max}\{\gamma_k\}$, where ${\rm max}\{ |U_{kp}|\}$ characterizes the strength of the long-range interaction between excitonic modes, and ${\rm max}\{\gamma_k\}$ the exciton decay rate. Assuming the polarization $\mathbf{P}(t)$ oscillates at two well-defined frequencies $\omega_c$ and $\omega_p$, we use the ansatz $\langle B_k(t)\rangle = X_k(t) {\rm e}^{-i\omega_pt}$, $\langle B_pB_q\rangle = Y_{pq}(t){\rm e}^{-i(\omega_p+\omega_c)t}$, $\langle a\rangle =A_c{\rm e}^{-i\omega_c t}$ and $\langle \mathcal{E}\rangle = {\rm e}^{-i\omega_p t}$ to separate eqs. (\ref{eq:B eom})-(\ref{eq:BB eom}) by frequency. The steady state solution for the probe susceptibility is $\epsilon_0\chi(\omega_p) =\sum_k\mu_kX_k/i\mathcal{E}_p$, with $\mu_k\equiv (\vec{\mu}_k\cdot \mathbf{e}_p)$. The one-exciton coherences $\mathbf{X} = [X_1,X_2,\ldots,X_N]^{\rm T}$ are obtained 
by solving the 
linear system
\begin{eqnarray}
 \mathbf{M}\mathbf{X} =  \mathbf{B}, 
 \label{eq:MX equation}
\end{eqnarray}
where $\mathbf{B} = [\mu_1,\mu_2,\ldots,\mu_N]^{\rm T}$ and 
\begin{eqnarray}
  \mathbf{M} = \left(\mathbf{O}   + 2|A_c|^2\;\mathbf{T}\right).
  \label{eq:M matrix}
\end{eqnarray}
The $N\times N$ one-photon detuning matrix is diagonal with elements $(\mathbf{O})_{nn} = i\Delta_{n}-\Gamma_{n} $, where $\Delta_n = \omega_c-\omega_{k_n}$ is the probe detuning from the $n$-th excitonic mode, and $\Gamma_n\equiv \Gamma_{k_n}$ is the decay rate. The coupling between the one-exciton band and the two-exciton band is accounted for in the two-photon detuning matrix $\mathbf{T}$, which has diagonal elements $(\mathbf{T})_{nn} = \sum_{j\neq n}^{N}D_{nj}^2/(i\Delta_{nj}-\Gamma_{nj})$ and off-diagonal elements $(\mathbf{T})_{mn} = D_{nm}D_{mn}/(i\Delta_{nm}-\Gamma_{nm})$. The elements $D_{nm}\equiv D_{k_n,k_nk_m}$ are proportional to the one-to-two exciton transition dipole matrix elements
\bibnote{In the bosonic approximation for the exciton states, we evaluate these matrix elements using the eigenstates $\ket{k}= \sum_ic_{ik}\ket{e_i}$ as $\langle k|\hat \mu|k q\rangle=\mu_{eg}\sum_{ij} c_{ik}^*c_{jk}c_{iq}+c_{ik}^*c_{ik}c_{jq}$, where we have defined the dipole operator in the site basis as $\hat \mu = \mu_{eg}\sum_m \ket{e_m}\bra{g_m}+\ket{g_m}\bra{e_m}$}.
The two-photon detuning is $\Delta_{nm} = \omega_p+\omega_c- \omega_{k_n}-\omega_{k_m}-2U_{nm}$. In the absence of the cavity we have $|A_c|=0$ and the linear response is simply given by a sum of Lorentzians centered at the exciton frequencies $\omega_k$, weighted by the 
corresponding transition dipole moments $g_k$. The coupling to the cavity therefore modifies the absorptive and dispersive response of the medium to the weak probe as described below. We note that the aggregate absorption spectra obtained from eq. (\ref{eq:MX equation}) satisfies the sum rule $\int{\rm Im}[\chi(\omega_p)]d\omega_p=N\pi/\epsilon_0$ for all values of $A_c$.

\begin{figure}[t]
 \includegraphics[width=0.50\textwidth]{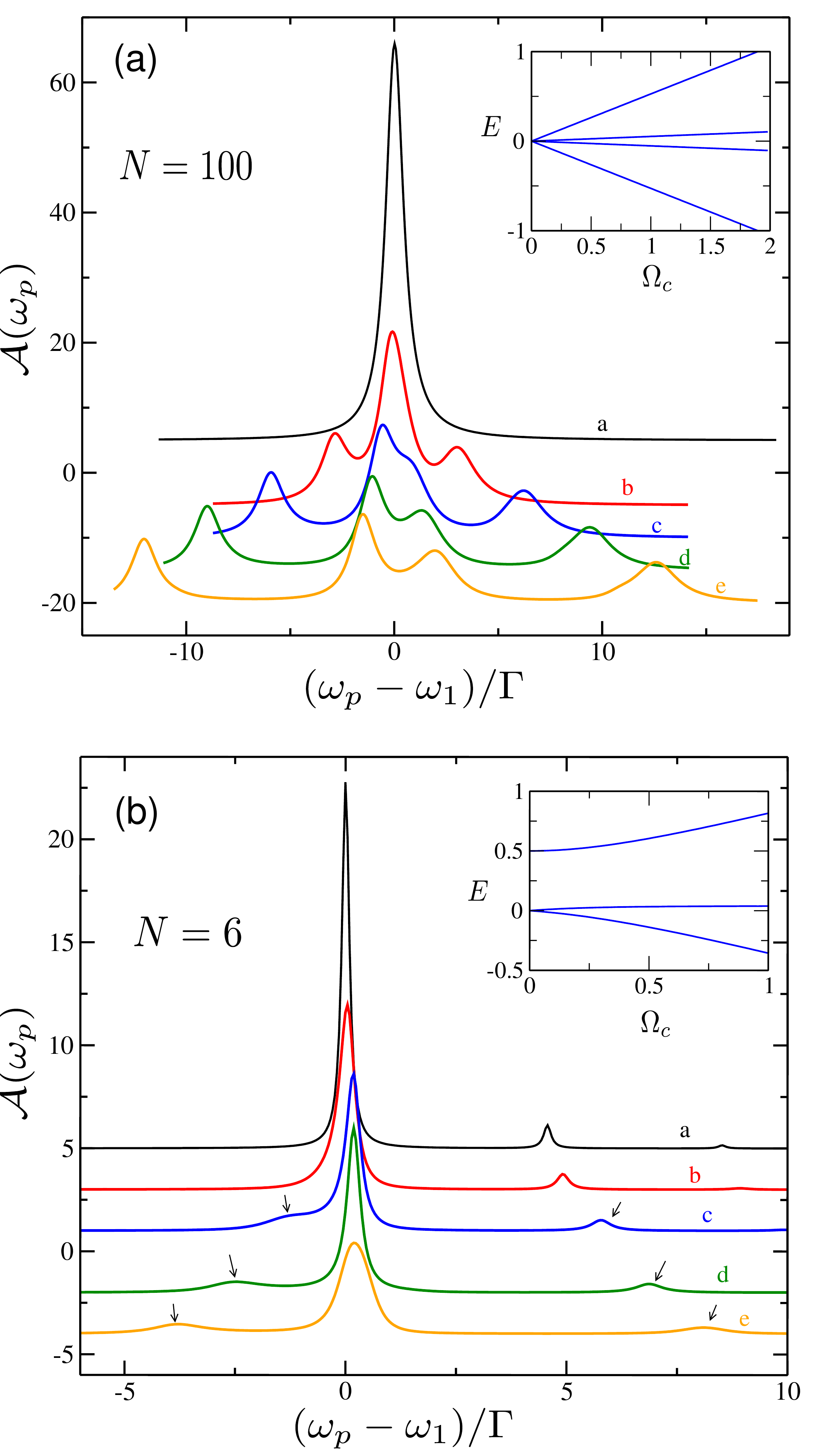}
 \caption{Probe absorption spectrum $\mathcal{A}(\omega_p)=-{\rm Im}[\chi]$ (in units of $\epsilon_0/\hbar$) as a function of the detuning from the lowest exciton resonance $\Delta_1=\omega_p-\omega_1$ (in units of the decay rate $\Gamma=26$ meV) for an ideal open linear polar J-aggregate of size $N$ in a microcavity (no energetic disorder). {\bf Panel a}: $N=100$. Curves are labeled according to the mean cavity amplitude $A_c\equiv|\langle a\rangle|$ as {\bf (a)} No cavity, {\bf (b)} $A_c= 0.2$, {\bf (c)} $A_c= 0.4$, {\bf (d)} $A_c= 0.6$, {\bf (e)} $A_c= 0.8$. 
 The inset shows the eigenvalues of the four-level effective Hamiltonian in eq. (\ref{eq:effective H4}) with $\Delta_2=0=\Delta_{23}$ and $\Omega_1=\Omega_2/3 = 3\Omega_3$ as a function of $\Omega_c=\Omega_2$ (in arbitrary units).
 {\bf Panel b}: $N=6$. Curves are labelled by the value of $A_c$ as {\bf (a)} No cavity, {\bf (b)} $A_c= 0.4$, {\bf (c)} $A_c= 0.8$, {\bf (d)} $A_c= 1.2$, {\bf (e)} $A_c= 1.6$. The inset shows the eigenvalues of the eq. (\ref{eq:effective H4}) with $\Delta_2=0.5$, $\Omega_1=\Omega_2/3$ and $\Omega_3=0$ as a function of $\Omega_c=\Omega_2$.
 In both panels the cavity frequency is resonant with the transition $\ket{k_1}\rightarrow\ket{k_1k_2}$. }
 \label{fig:1D aggregate}
\end{figure}

In order to illustrate the developed model in eq. (\ref{eq:MX equation}), we calculate the probe susceptibility $\chi(\omega_p)$ for open one-dimensional homogeneous aggregates of size $N$. In Fig. \ref{fig:1D aggregate} we show the computed absorption probe spectra for PIC J-aggregates with $N=100$ and $N=6$. We use the transition energy $\omega_e = 2.25$ eV for all sites, nearest-neighbour excitonic coupling $J_{12} = -68.2$ meV, and dipole-dipole coupling $U_{12} = -198$ meV. These parameters were obtained by Markov {\it et al.}\cite{Markov:2000} from the observation of a red-shifted induced absorption peak in the pump-probe spectra of PIC aggregates. The relaxation tensors $\Gamma_k$ and $\Gamma_{pq}$ are dominated by phonon scattering (see SI), and for simplicity we set $\Gamma_k=\Gamma_{pq}\approx 26$ meV, which gives a lower bound for exciton dephasing rate at room temperature\cite{Valleau:2012}. The vacuum 
Rabi frequency in microcavities 
can reach values on the order of $\Omega_c \sim 100$ meV~\cite{Lidzey:1998}. 
In Fig. \ref{fig:1D aggregate} we set the vacuum Rabi frequency $\Omega_c\equiv \sqrt{N_A}\bra{e}\vec{\mu}\cdot \mathbf{e}_c\ket{g}\mathcal{E}_c =\Gamma_k$ so that $D_{k,kp}\sim \sqrt{N}\Omega_c$ for the strongest excitonic transitions exceeds the dissipation rates, as is required in the strong coupling regime. 

The probe absorption spectra in Fig. \ref{fig:1D aggregate}a displays a four-peak structure where the peak splitting scales linearly with the mean cavity amplitude $\langle a\rangle $. The free-space spectrum corresponds to the J-band. This trend can be qualitatively explained using a semiclassical model in which a classical cavity field of frequency $\omega_c$ couples strongly with two states in the one-exciton band (labelled $\ket{k_1}$ and $\ket{k_2}$) and two states in the two-exciton band ($\ket{k_1k_2}$ and $\ket{k_2k_3}$). The coupling scheme is illustrated in Fig. 1c. The transition dipole moments from the ground state $\ket{g}$ to the states $\ket{k_1}$, $\ket{k_2}$ and $\ket{k_3}$ have the largest values in the one-exciton band and satisfy $\mu_1>\mu_2>\mu_3$. The cavity frequency is chosen to be on resonance with the $\ket{k_1}\rightarrow\ket{k_1k_2}$ transition. The effective Hamiltonian $\mathcal{H}_{\rm eff}$ that describes the couplings between the two bands in the rotating frame of the cavity field 
is given by 
\begin{eqnarray}
 \mathcal{H}_{\rm eff} = \left(\begin{array}{cccc}
                      0 & \Omega_1& 0& 0\\
                      \Omega_1 & \omega_{12}-\omega_c & \Omega_2&0 \\
                      0&\Omega_2&\Delta_2 & \Omega_3\\
                      0&0&\Omega_3&\omega_{12}+ \Delta_{23} -\omega_c
                     \end{array}
\right),
\label{eq:effective H4}
\end{eqnarray}
where the frequency parameters are defined in Fig. 1c. Energy is measured with respect to the lowest exciton state $\ket{k_1}$, i.e., $\omega_1\equiv0$. For a large homogeneous aggregate, the excitonic bands become quasi-continuous and the splittings $\Delta_2 \equiv \omega_{2}-\omega_{1}$ and $\Delta_{23} \equiv \omega_{23}-\omega_{12}$ become negligibly small in comparison with typical linewidths. Therefore we can assume that the cavity strongly couples almost on resonance the four levels shown in Fig. 1c. In this regime, a weak probe field will drive transitions between the ground state $\ket{g}$ and the eigenvalues of the effective Hamiltonian $\mathcal{H}_{\rm eff}$ in eq. (\ref{eq:effective H4}), which are the new normal modes of the cavity-matter system. In the inset of Fig. \ref{fig:1D aggregate}a we show the eigenvalues of $\mathcal{H}_{\rm eff}$ for $\omega_c=\omega_{12}$, $\Delta_2=\Delta_{23}=0$, $\Omega_2=3\Omega_1$ and $\Omega_3=\Omega_2/3$ as a function of $\Omega_c\equiv\Omega_2$. For $\Omega_
c\ll 
1$ (in arbitrary units) only three peaks can be resolved, but as $\Omega_c$ increases the middle peak splits into a doublet, giving rise to the four-peak structure observed in the probe spectra calculated using eq. (\ref{eq:MX equation}), which includes $N$ one-exciton states and $\sim N^2$ two-exciton states.

The number of states in the one and two-exciton bands that couple strongly with the cavity field depends on the size of the molecular aggregate. In order to illustrate this fact, we show in Fig. \ref{fig:1D aggregate}b the probe absorption spectrum for an open 1D homogeneous aggregate of size $N=6$, with the same values of $\omega_e$, $J_{12}$ and $U_{12}$ and decay parameters as in Fig. \ref{fig:1D aggregate}a. The cavity field is again resonant with the $\ket{k_1}\rightarrow\ket{k_1k_2}$ transition, but now the states $\ket{k_2}$ and $\ket{k_3}$ are no longer quasi-degenerate with $\ket{k_1}$ because of the small array size. The cavity frequency is thus detuned from their corresponding transitions with the states $\ket{k_1k_2}$ and $\ket{k_2k_3}$ in the two-exciton band. Since the Rabi frequency $\Omega_3$ is proportional to the transition dipole $\mu_{k_2,k_2k_3}$ by construction, whenever $\Omega_3/\Delta_{23}\ll 1$, we can set $\Omega_3=0$ in eq. \ref{eq:effective H4} to effectively remove the state $\ket{k_2k_3}$ from the excited state dynamics. Interestingly, the eigenstates of the resulting three-level system with $\Delta_2>0$, plotted as a function of $\Omega_c$ in the inset of Fig. \ref{fig:1D aggregate}b, show a trend in very good agreement with the microscopic model derived in eq. (\ref{eq:MX equation}).

\begin{figure}[t]
 \includegraphics[width=0.5\textwidth]{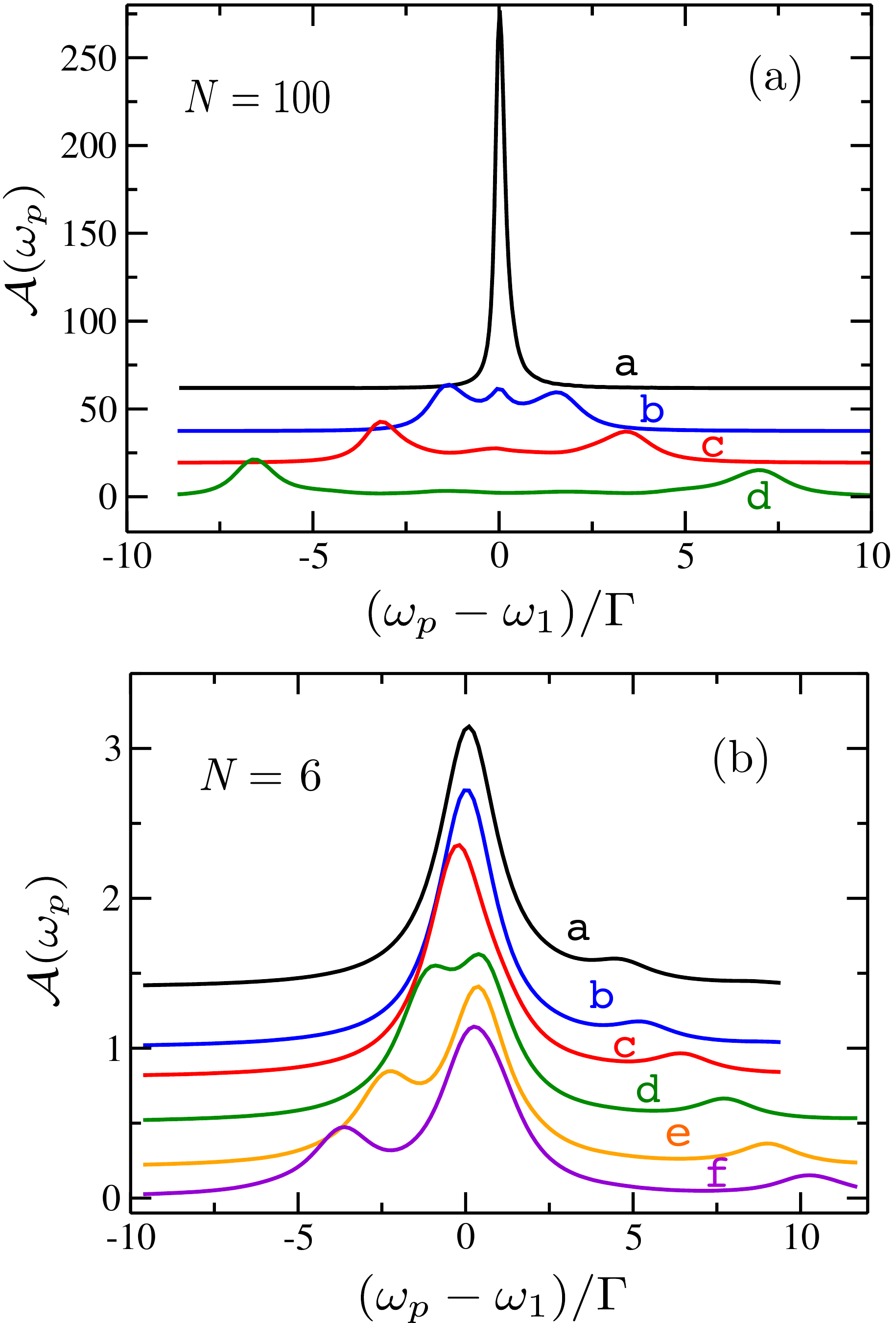}
 \caption{Probe absorption $\mathcal{A}(\omega_p)$ for a 1D J-aggregate as a function of the probe detuning from an arbitrarily chosen one-exciton state near the bottom of the one-exciton band.   Dipolar couplings $J_{ij}$ and $U_{ij}$ are the same as in Fig. \ref{fig:1D aggregate}a. 
  The cavity frequency is chosen such that the two-photon detuning vanishes when $\omega_p-\omega_1=0$. {\bf Panel a}: Aggregate size is $N=100$. Curves are labeled according to the mean cavity amplitudes: {\bf (a)} No cavity, {\bf (b)} $A_c =|\langle a\rangle|= 0.1$, {\bf (c)} $A_c= 0.2$, {\bf (d)} $A_c= 0.3$. {\bf Panel b}: Aggregate size is $N=6$ for {\bf (a)} No cavity, {\bf (b)} $A_c = 0.4$, {\bf (c)} $A_c= 0.8$, {\bf (d)} $A_c= 1.2$, {\bf (e)} $A_c= 1.6$, {\bf (f)} $A_c = 2.0$. In both panels the vacuum Rabi frequency is $\Omega_c = \Gamma$, where $\Gamma=26 $ meV is the exciton decay rate. Static disorder is modelled by taking each monomer energy randomly from a Gaussian distribution with mean $E_0=2.25$ eV and standard deviation $\sigma = 0.125 J$. 
a wide frequency range for a low number of cavity photons. $\gamma_e$ is the single-molecule gas-phase radiative decay rate.
}
 \label{fig:inhomogenous}
\end{figure}

We now include the effect of inhomogeneous disorder in the evaluation of the probe absorption spectra. In order to model static energetic disorder we assume the site energies are given by $\varepsilon_i=\omega_e+d_i$, where  $d_i$ is a random energy shift taken independently for each site from a Gaussian distribution with standard deviation $\sigma/|J|\sim 0.1$, consistent with the motional narrowing limit \cite{Knapp:1984,Knoester:1993}. The susceptibility $\chi_p$ obtained from eq. (\ref{eq:MX equation}) needs to be averaged over the ensemble of disorder realizations. 
In Fig. \ref{fig:inhomogenous} we show the probe absorption spectrum $\mathcal{A}(\omega_p)$ of the same J-aggregates used in Fig. \ref{fig:1D aggregate}a with $N=100$ and $N=6$ molecules, but now static disorder is introduced in the Hamiltonian. The structure of the spectrum resembles the results in Fig. \ref{fig:1D aggregate}a, but the details of the splittings near the origing are different because now the detunings $\Delta_2$ and $\Delta_{23}$ in eq. (\ref{eq:effective H4}) are now averaged over an ensemble of disorder realizations. The behaviour of the outer peaks persist in the presence of disorder as a result of the strong interaction between the cavity mode and the transition $\ket{k_1}\rightarrow\ket{k_1k_2}$, close to the deterministic resonance frequency $\omega_{12} = \omega_2-2|U_{12}|$. 

\begin{figure}[t]
 \includegraphics[width=0.48\textwidth]{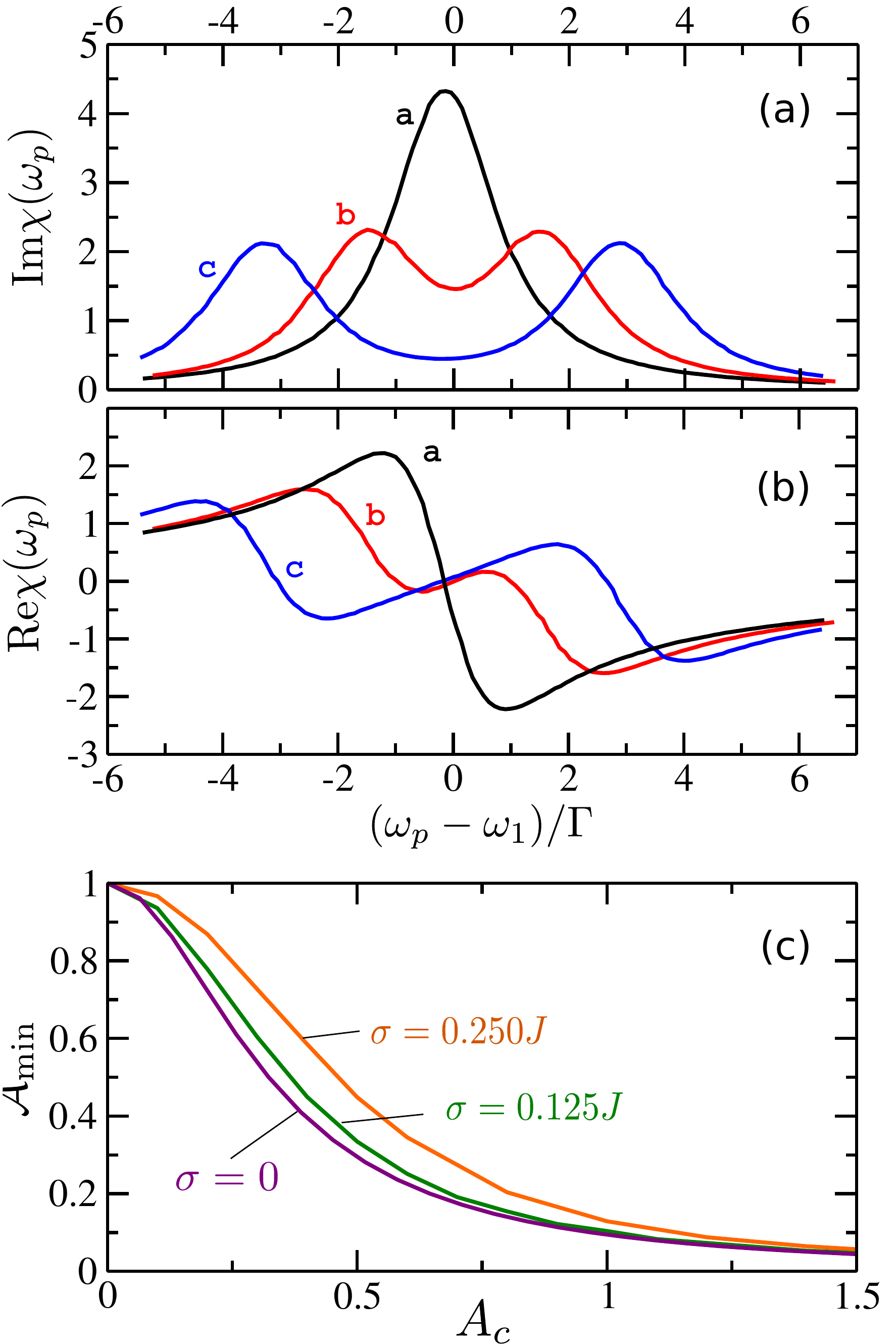}
 \caption{Im [$\chi_p$] ({\bf panel a}) and Re[$\chi_p$] ({\bf panel b}) for an inhomogeneously broadened dimer as a function of the probe detuning from the maximum of free-space exciton absorption band. 
 Dipolar couplings $J_{ij}$ and $U_{ij}$ are the same as in Fig. \ref{fig:1D aggregate}. 
 The cavity frequency is chosen such that the two-photon detuning vanishes when $\omega_p-\omega_1=0$.
 Curves are labeled according to the mean photon number $A_c=|\langle a\rangle|$: {\bf (a)} No cavity, {\bf (b)} $A_c= 0.5$, {\bf (c)}$A_c= 1.0$.
 The vacuum Rabi frequency is $\Omega_c = 5 \Gamma$, where $\Gamma = 26$ meV is the exciton decay rate. {\bf Panel c}:
 Minimum absorption $\mathcal{A}_{\rm min}$ (normalized to the free-space value) near the deterministic resonance $\omega_p-\omega_1=0$ (between the Autler-Townes doublet), as a function of the mean cavity amplitude $A_c$. Curves are labelled by the width of the Gaussian distribution of static energy shifts $\sigma$ (in units of the exchange dipole coupling $J$). All other parameters are the same as in panels a and b.}
 \label{fig:dimer}
\end{figure}

As a final example, we consider the response of a dimer of coupled polar chromophores~\cite{Halpin:2014}. Clearly for $N=2$ the bosonic approximation used in the derivation of eq. (\ref{eq:MX equation}) is no longer valid to describe the two-exciton manifold, which now consists of a single state $\ket{e_1e_2}$. However, since the structure of eqs. (5) and (6) is universal, the steady-state solution $\mathbf{X}$ in eq. (\ref{eq:MX equation}) remains valid by simply redefining the elements of the two-photon detuning matrix $\mathbf{T}$. The one-exciton manifold has states $\ket{\psi_+} = \sqrt{a}\ket{e_1g_2}+\sqrt{1-a}\ket{g_1e_2}$ and $\ket{\psi_-} = -\sqrt{1-a}\ket{e_1g_2}+\sqrt{a}\ket{g_1e_2}$ with $0\leq a\leq 1$. The transition dipole moments from the one-exciton manifold to the ground and two-exciton states are given by $\mu_{\pm}\equiv\bra{\psi_\pm}\vec{\mu}\ket{g_1g_2} =\mu_{eg}\left(\sqrt{a}\pm\sqrt{1-a}\right)= \bra{\psi_\pm}\vec{\mu}\ket{e_1e_2}$. Given that the two-exciton energy  $\omega_{12} =\
epsilon_1+\epsilon_2-|U_{12}|$ is red-shifted with respect to the one-exciton transition frequencies $\omega_\pm$, the response of a single dimer to a probe field at frequency $\omega_p$, when the cavity is resonant with the $\ket{\psi_+}\rightarrow\ket{e_1e_2}$ transition, is given by 
\begin{eqnarray}\label{eq:TMA}
 \chi(\omega_p) = i\left[\frac{\mu_1^2}{\epsilon_0}\right]\frac{(\Gamma_{12}-i[\Delta_p+\Delta_c])}{(i\Delta_p - \Gamma_1)(i[\Delta_p+\Delta_c]-\Gamma_{12})+2D_{12}^2|A_c|^2},
\end{eqnarray}
which corresponds to eq. (\ref{eq:MX equation}) in the limit where $\mathbf{M}$ has a single non-zero element. Here $\Delta_p = \omega_p-\omega_+$, $\Delta_c = \omega_c-(\epsilon_1+\epsilon_2+U_{12}-\omega_+)$, $\mu_1 =\mu_+$ and $D_{12}\propto \mu_+$. $\Gamma_1$ and $\Gamma_{12}$ are the one and two-exciton decoherence rates. 
Equation (\ref{eq:TMA}) shows that for coupled dimers the cavity acts as a control for the propagation of the weak probe, in analogy with the phenomenology stemming from atomic physics to describe electromagnetically-induced transparency (EIT) in a cascaded three-level system \cite{Fleischhauer:2005}. Figure \ref{fig:dimer} shows the absorptive and dispersive response of an inhomogeneously broadened dimer of polar chromophores in panels a and b, respectively. The probe susceptibility has the standard features of EIT: reduction of probe absorption and steep dispersion on resonance with the lowest exciton state\cite{Fleischhauer:2005}. In comparison with atomic systems, the EIT linewidth for the dimer is broader even in the absence of static disorder because the two-exciton coherence is short-lived, i.e., $\Gamma_{12}/\Gamma_1\sim 1$. Inhomogeneous broadening further broadens the EIT features, and in particular increases the absorption minimum under conditions of one and two-photon resonances $\Delta_p=0=\Delta_c$. 
For homogeneously broadened dimers (described by eq. (\ref{eq:TMA})), the absorption minimum for a resonant probe is plotted in Fig. \ref{fig:dimer}c, showing a scaling of  $\mathcal{A}_{\rm min}\approx (\Gamma_{12}/2D_{12}^2)A_c^{-2}$ for large cavity coupling $D_{12}A_c\gg \Gamma_1\sim \Gamma_{12}$. The homogeneous curve gives a lower bound for the probe absorption minimum on resonance. 
The strength of the static disorder in the site basis is given by $\sigma$ as before. We average over an ensemble of 1200 realization of the site energy shifts $(d_1,d_2)$ using an uncorrelated Gaussian joint probability distribution (JPD) of the form $P(d_1,d_2) = P(d_1)P(d_2)$. 
Increasing the disorder strength $\sigma$, increases the resonant absorption of the probe $\mathcal{A}_{\rm min}$ for intermediate values of $D_{12}A_c$. However, as the strength of the cavity coupling increases the homogeneous limit is recovered. This behaviour has already been observed for EIT in Doppler-broadened atomic gases\cite{Gea-Banacloche:1995}. 

In order to gain qualitative analytic understanding of the absorption minimum $\mathcal{A}_{\rm min}$ for inhomogeneously broadened dimers with $\sigma\neq 0$, we evaluate the mean susceptibility $\langle \chi(\omega_p)\rangle$ directly from eq. (\ref{eq:TMA}) by averaging over an ensemble of one- and two-exciton detunings $\Delta_p=\Delta_p^{(0)}-D_p$ and $\Delta_c=\Delta_c^{(0)}-D_c$, where the random shifts $(D_p,D_c)$ ultimately result from the site energetic disorder $(d_1,d_2)$. We integrate over all possible shifts using $\langle \chi(\omega_p)\rangle  = \int\int dD_pdD_c \,\chi(\omega_p,D_p,D_c)P(D_p,D_c)$, where $P(D_p,D_c)$ is the JPD for the one- and two-exciton shifts. In general, $P(D_p,D_c)$ does not factorize even if $P(d_1,d_2)$ does, due to F\"orster coupling $J_{ij}$\cite{Knoester:1993}. In order to simplify the integration over disorder, we assume an uncorrelated JPD of the form $P(D_p,D_c) = P(D_p)P(D_c)$, where $P(D) = \pi^{-2} \gamma /\left(\gamma^2+{D}^2\right)$ is the Cauchy distribution with width $\gamma$. The probe absorption under conditions of deterministic one- and two-photon resonances $\Delta^0_p=0=\Delta_c^0$ thus gives
\begin{eqnarray}\label{eq:Amin disorder}
\mathcal{A}_{\rm min}=\frac{ \Gamma_{12}+\gamma_p+\gamma_c}{(\Gamma_1+\gamma_p) (\Gamma_{12}+\gamma_p+\gamma_c)+2D_{12}^2A_c^2}, 
\end{eqnarray}
where $\gamma_p$ and $\gamma_c$ are the widths of $P(D_p)$ and $P(D_c)$, respectively. We find that $\mathcal{A}_{\rm min}$ in eq. (\ref{eq:Amin disorder}) for a Cauchy distribution provides an upper bound for the results obtained by numerically averaging the independent site disorder over an Gaussian distribution with the same width. However, in the limit $A_c\gg (\Gamma_{1}+\gamma_p)/D_{12}$, eq. (\ref{eq:Amin disorder}) gives $\mathcal{A}_{\rm min}\approx [(\Gamma_{12}+\gamma_p+\gamma_c)/2D_{12}^2]A_c^{-2}$, which tends towards the homogeneous limit for short-lived two-exciton coherences $\Gamma_{12}\gg (\gamma_p+\gamma_c)$, which is the case considered here for dimers.

In summary, we present in this Letter a general scheme to perform nonlinear optical experiments using polar J-aggregates at the single-photon level. The setup involves the use of organic chromophores with a moderate to large difference between ground and excited state permanent dipole moments $\Delta d$, that can assemble into low-dimensional aggregate structures. We have illustrated our findings using pseudoisocyanine chloride (PIC) dyes, but the conclusions of this work are general. Organic chromophores with large $\Delta d\sim 1-10 $ Debye continue to be under active experimental investigation for the design of second-order nonlinear optical materials~\cite{Verbiest:1997,Luo:2009,Li:2013}. Upon aggregation, these polar dyes can lead to strong exciton-exciton interactions that exceed the broadening of the exciton line. For attractive interactions (J-aggregation), the cavity field can be used to strongly drive coherences between the one- and two-exciton bands without removing population from the ground state of an aggregate. 
Under these conditions the absorption of a weak probe field resonant with the cavity-free exciton absorption peak is significantly modified by the presence of the cavity field containing a single photon (on average), which can be seen as quantum optical switching. In order to achieve this effect it is important that the cavity-matter coupling exceeds all the dissipation rates in the system, a regime that is experimentally accessible~\cite{Lidzey:1998,Kena-Cohen:2010}. We have restricted our discussion to strong light-matter coupling in optical microcavities, but the strong coupling regime has also been achieved for molecular aggregates in the near-field of plasmonic nanostructures \cite{Bellessa:2004,Wurtz:2007,Vasa2013}, which further opens the applicability of our proposed scheme to sub-wavelength nonlinear quantum optics. 

The ability to control molecular aggregates in optical nanostructures not only offers opportunities for the development of novel organic-based optical devices~\cite{Saikin:2013}, but we envision new possibilities of quantum control of excited state dynamics relevant in energy transport and chemical reactivity, and engineering of excitonic materials that are topologically robust against disorder \cite{Yuen2014}. Current experiments can achieve the regime of ultrastrong coupling with organic ensembles, where the light-matter interaction strength can be a significant fraction of the chemical binding energy\cite{Schwartz:2011}. In this regime, it should be possible to control the outcome of chemical reactions at the level of thermodynamics by effectively lowering reaction barriers~\cite{Hutchison:2012}, in analogy with traditional catalytic processes, thus directly affecting reaction kinetics. This novel strong-field single-photon quantum control paradigm for molecular processes should be contrasted with traditional strong-field laser control schemes that require very high laser intensities to modify the chemical energy landscape~\cite{Corrales:2014}, or weak-field coherent control schemes that exploit delicate laser-induced quantum interferences among internal vibronic states~\cite{Shapiro-Brumer-book}, but do not modify the energetics of the reaction. Quantum optical control of chemical dynamics is a future research direction with promising applications in nanoscience and technology, where traditional bulk methods for controlling chemical reactivity have limited efficiency.



\begin{acknowledgement}
We thank Frank Spano and Thibault Peyronel for discussions. F.H. and A.A.-G. acknowledge the support from the Center for Excitonics, an Energy Frontier Research Center funded by the U.S. Department of Energy, Office of Science and Office of Basic Energy Sciences, under Award Number DE-SC0001088. 
F.H., S.K.S. and A.A.G. also thank the Defense Threat Reduction Agency Grant HDTRA1-10-1-0046. 
S.K.S. is also grateful to the Russian Government Program of Competitive Growth of Kazan Federal University. L.A.P. acknowledges support from the \emph{Comit\'e para el Desarrollo de la Investigaci\'on} -CODI-- of Universidad de Antioquia, Colombia under the Estrategia de Sostenibilidad 2014-2015, and by the \emph{Departamento Administrativo de Ciencia, Tecnolog\'ia e Innovaci\'on} --COLCIENCIAS--  of Colombia under the grant number 111556934912.
\end{acknowledgement}

\begin{suppinfo}
The derivation of the quantum Langevin equations leading to eqs. (\ref{eq:B eom})-(\ref{eq:BB eom}), and the derivation of the effective Hamiltonian in eq. (\ref{eq:effective H4}) can be found in the Supporting Information (SI). 

\end{suppinfo}

\clearpage
\bibliography{JaggregateEIT}

\begin{mcitethebibliography}{48}
\bibitem[Not()]{Note-1}
In the bosonic approximation for the exciton states, we evaluate these matrix
  elements using the eigenstates $\ket{k}= \sum_ic_{ik}\ket{e_i}$ as $\langle
  k|\hat \mu|k q\rangle=\mu_{eg}\sum_{ij}
  c_{ik}^*c_{jk}c_{iq}+c_{ik}^*c_{ik}c_{jq}$, where we have defined the dipole
  operator in the site basis as $\hat \mu = \mu_{eg}\sum_m
  \ket{e_m}\bra{g_m}+\ket{g_m}\bra{e_m}$\relax
\mciteBstWouldAddEndPuncttrue
\mciteSetBstMidEndSepPunct{\mcitedefaultmidpunct}
{\mcitedefaultendpunct}{\mcitedefaultseppunct}\relax
\EndOfBibitem
\end{mcitethebibliography}

\end{document}